\documentclass[aps, prd, twocolumn, preprintnumbers, superscriptaddress, nofootinbib, floatfix]{revtex4-2}

\usepackage{amsmath}
\usepackage{bm}
\usepackage{amsfonts}
\usepackage{graphicx}
\usepackage{epstopdf}
\usepackage{hyperref}
\usepackage{array}
\usepackage[margin=1in,footskip=0.2in]{geometry}
\usepackage{soul}
\usepackage{color}

\enlargethispage{\baselineskip}

\hypersetup{
colorlinks=true,
linkcolor=blue,
linktoc=page,
citecolor=blue,
urlcolor=blue}

\begin{document}
\title{B-field induced mixing between Langmuir waves and axions}

\author{Luca Visinelli}
\email[Electronic address: ]{luca.visinelli@sjtu.edu.cn}
\affiliation{Tsung-Dao Lee Institute (TDLI), 520 Shengrong Road, 201210 Shanghai, P.\ R.\ China}
\affiliation{School of Physics and Astronomy, Shanghai Jiao Tong University,\\ 800 Dongchuan Road, 200240 Shanghai, P.\ R.\ China}

\author{Hugo Ter\c{c}as}
\email[Electronic address: ]{hugo.tercas@tecnico.ulisboa.pt}
\affiliation{Instituto de Plasmas e Fus\~ao Nuclear, Av.\ Rovisco Pais 1, 1049-001 Lisboa, Portugal}
\affiliation{Instituto Superior T\'ecnico, Av.\ Rovisco Pais 1, 1049-001 Lisboa, Portugal}

\date{\today}
\preprint{NORDITA-2018-050}

\begin{abstract}
We present an analytic study of the dispersion relation for an isotropic magnetized plasma interacting with axions. We provide a quantitative picture of the electromagnetic plasma oscillations in both the ultrarelativistic and nonrelativistic regimes and considering both non-degenerate and degenerate media, accounting for the dispersion curves as a function of the plasma temperature and the ratio of the plasma phase velocity to the characteristic velocity of particles. We include the modifications on the Landau damping of plasma waves induced by the presence of the axion field, and we comment on the effects of damping on subluminal plasma oscillations.
\end{abstract}

\maketitle

\section{Introduction}

The study of the physics of plasmas is crucial for the understanding of a vast array of applications ranging from laboratory experiments to astrophysics and cosmology. To some extents, a fraction of the interstellar medium and the intergalactic space, the interior of stars, the solar wind, and accretion discs can be treated using plasma physics. Examples of high-energy density plasmas are the interior of giant planets~\cite{Chabrier:2009} and white dwarfs~\cite{Chabrier:2002hx}, the atmosphere of neutron stars~\cite{Harding:2006qn}, and the interaction of plasmas with petawatt lasers~\cite{PhysRevLett.83.4709, PhysRevLett.98.065002}.

Any external perturbation, such as an incident electromagnetic wave, may change the equilibrium of the plasma and drive both electrostatic and electromagnetic oscillations~\cite{PhysRev.33.195, PhysRev.34.876}. Oscillations in isotropic plasmas are of electrostatic (longitudinal, also known as {\it Langmuir oscillations}) and electromagnetic (transverse) characters. The corresponding dispersion relations have been derived for both nonrelativistic plasmas~\cite{Thomson:1934, Vlasov:1938, PhysRev.75.1851, PhysRev.75.1864} as well as for ultrarelativistic plasmas~\cite{Clemmow117, Silin:1960, Silin:1960pya, Buti:1962, Braaten:1993jw, Dettmann:1993zz}. The quantum of plasma oscillations, the plasmon~\cite{Pines:1952zz, Bohm:1953zza}, might couple strongly with an external electromagnetic field to form mixed states called polaritons~\cite{tolpygo:1950}.

Plasmon dynamics could be also altered by their interactions with light fields, thus serving as efficient detectors for the presence of new physics~\cite{Ostman:2004eh, Lai:2006af, Ganguly:2008kh, Dvorkin:2019zdi}. One such example is the hypothetical axion~\cite{Weinberg:1977ma, Wilczek:1977pj}, a light pseudoscalar field originally introduced as a solution to the strong CP problem within the QCD sector of the standard model (see Refs.~\cite{Marsh:2015xka, DiLuzio:2020wdo} for recent reviews). Outside of the QCD theory, light axions might arise copiously in string theory embeddings~\cite{Choi:2006qj, Svrcek:2006yi}, possibly leading to a plethora of light fields in the string ``axiverse''~\cite{Arvanitaki:2009fg, Visinelli:2018utg} and to alternative explanations for the DM puzzle~\cite{Arias:2012az, Visinelli:2017imh}. Coupling an axion with electrically charged fermions and with photons introduces new Lagrangian terms that modify the action for the electromagnetic field and leads to new effects in electrodynamics~\cite{Kaplan:1985dv, Srednicki:1985xd, Sikivie:1983ip, Wilczek:1987mv, Visinelli:2013fia}. Several experiments, either proposed or already deployed, aim to scan the possible values of the axion mass window and coupling with photons~\cite{Turner:1989vc}. For recent reviews of some experimental techniques and active searches see Refs.~\cite{Irastorza:2018dyq, Sikivie:2020zpn}.

The immediate interest in axion-plasmon polaritons is two-fold: first, polaritons can be used to investigate the effects of plasmas in both astrophysical and cosmological phenomena, impacting on the properties of the cosmic microwave background~\citep{Gariazzo:2017pzb}, the primordial axion abundance problem~\citep{Raffelt:2006cw, Kamada:2017cpk}, and the production of axions in the stellar medium~\citep{Bernstein:1988bw, Dominguez:2017mia, Meyer:2016wrm}. Second, polaritons can be used as a direct signature of axions in the laboratory, where the different experiments to be designed in the near future and the corresponding sensitivities can be tuned by changing the density of the electrons in the plasmas. As such, discharge, tokamaks, solid-state and ultracold plasmas emerge as potential platforms for polariton experiments. Recently, plasma metamaterials have proposed to increase the sensitivity of axion haloscopes towards the QCD parameter range \cite{Lawson:2019}. 

Astrophysical regions possessing a strong magnetic field have been proposed as laboratories to search for axions. As an example, the synchrotron emission of axions interacting with the magnetosphere surrounding pulsars would affect the luminosity of neutron stars both through an axion-electron coupling at tree-level~\cite{Borisov:1994wg, Kachelriess:1997kn} and the mediation of a plasmon~\cite{Mikheev:1998bg, Mikheev:2009zz, Caputo:2020quz}. Similarly, axion-photon interactions in the magnetosphere of neutron stars could lead to detectable radio signals~\cite{Pshirkov:2007st, Huang:2018lxq, Hook:2018iia, Safdi:2018oeu, Edwards:2019tzf, Edwards:2020afl}. Screening effects in relativistic plasmas and the production of thermal axions have also been considered, and the estimate bounds the coupling coupling constant between axions and photons based on data from the supernova SN1987A event~\cite{Altherr:1990wi, Altherr:1990tf, Braaten:1991dd, Salvio:2013iaa}. Note, that a similar search is currently undergoing for hypothetical ``dark'' photons, whose mixing with Langmuir waves could alter stellar evolution~\cite{An:2013yfc, Redondo:2013lna, Bernal:2017kxu}.

As for laboratory experiments, axions in plasmas have recently gained attention in the context of intense lasers. In fact, the next generation of high-power laser facilities is expected to provide conditions to probe QED physics in parameter regimes that are inaccessible to particle colliders~\cite{DiPiazza:2011tq}. The Extreme Light Infrastructure (ELI) experiment, for example, will offer the possibility to investigate the effect of the Heisenberg-Euler vacuum (virtual electron-positron pairs)~\citep{DiPiazza:2011tq, Popovici_2017} and the quantum recoil due to radiation emission~\citep{Burton:2014wsa}. The wakefield acceleration paradigm gained much breath as it reveals to be an efficient way to accelerate particles~\citep{Mangles:2004ta, Geddes:2004tb, Faure:2004tc}, and recent studies have exploited such wakefields~\cite{Tajima:1979bn} to produce axions in the lab~\citep{Mendonca:2007zz, Burton:2017bxi, Burton:2009bt, Burton:2015qda}. These laser-plasma interaction facilities differ from the setups discussed above, in the sense that axions could in principle be produced \textit{in situ}, with their presence manifesting in the features of the plasma.

In previous work, one of us has shown that in a magnetized plasma, an electromagnetic field is modified by the presence of the axion field through its dynamical backreaction mediated by the Lorentz force~\cite{Tercas:2018gxv, Mendonca:2019eke}. This results in a non-resistive, ``dielectric'' coupling between the axion and the plasmon, which leads to the excitation of a novel quasiparticle, the axion-plasmon polariton. As such, we expect this novel axion-plasmon polariton to pave the stage for a new research branch in the physics of axions.

In this paper, we establish a generic kinetic theory of axion-plasmon polaritons. Our approach is based on a phase-space description of the axion-plasmon coupling, going beyond the hydrodynamical treatment presented in Ref.~\citep{Tercas:2018gxv}. The theoretical framework is based on the Vlasov equation for a relativistic plasma~\citep{Silin:1960pya, Dettmann:1993zz}, which is coupled, via the modified Maxwell's equations, to the Klein-Gordon equation describing the axion. Such a theoretical framework is able to capture effects that are not cast by the hydrodynamics formulation, such as wave-wave and wave-particle interaction, and the isentropic damping of the collective plasma oscillations in the presence of hot electrons, the so-called \textit{Landau damping}~\citep{LIFSHITZ1981217, Chen:book}. The isentropic, energy conserving nature of the Landau damping is a particular feature of the kinetic formulation, being completely absent in hydrodynamical or single-particle models of plasmas~\cite{landau:book5, 2000JPlPh..63..371B}. As such, a theoretical theory of axions in a plasma opens the venue for new phenomena in astrophysical and cosmological scenarios that have so far remained elusive. The axion-photon coupling modifying the plasmon dispersion relation has been explored in previous work~\cite{Mikheev:1998bg, Mikheev:2009zz, Caputo:2020quz} to compute the emission rate of the process $e^- \to e^- + a$ mediated by a plasmon enhanced at axion-plasmon crossing, in relation with the energy loss from a supernova core and to predict the spectrum of axions radiated from the Sun. Here, we derive the condition for crossing in a general setup, including that of a relativistic plasma which has not been yet covered in the literature.

This paper is organized as follows. In Sec.~\ref{sec:maxwell}, we revise the system combining Maxwell equations with the Klein-Gordon equation. In Sec.~\ref{dispersion}, we derive the kinetic dispersion relation for the axion-plasmon polariton. In Sec.~\ref{results}, we specialize our results to a positron-electron plasma, for which we derive the dispersion relation, and we discuss analytically its features in various scenarios, namely for the case of an ultrarelativistic, degenerate, and nonrelativistic plasmas. We provide the numerical calculations for the ultrarelativistic case, which is unexplored in the literature. Discussion and conclusions are drawn in Sec.~\ref{conclusion}. We work in units $c = \hbar = 1$.

\section{Maxwell equations coupled to an axion field}
\label{sec:maxwell}

Whenever the axion field $\varphi$ couples with electrically-charged fermions $\psi$, an effective axion-photon coupling $g_{\varphi\gamma} \equiv \alpha_{\rm EM}\,\xi/(2\pi f)$ arises through fermion loop diagrams, where $\alpha_{\rm EM}\equiv e^2/(4\pi)$ is the fine-structure constant, $\xi$ is a model-dependent parameter, and $f$ is the axion decay constant. In the case of the QCD axion, the mixing with pions due to the interaction with the gluons leads to a coupling whose numerical value depends on the electromagnetic and color anomalies of the embedding theory~\cite{Kaplan:1985dv, Srednicki:1985xd}. More generally, $g_{\varphi\gamma}$ parametrizes the strength of the coupling for an axion interacting with a photon field. The axion also couples at tree level with the fermion current of a Dirac fermion field, with the coupling strength $g_{\varphi\psi}$.

In this setting, the Lagrangian density that describes the electromagnetic field $A_\mu$ interacting with a Dirac fermion $\psi$ of mass $m_\psi$ and charge $q$, and an axion $\varphi$ of mass $m_\varphi$ reads
\begin{eqnarray}
	 \label{eq:lagrangian}
	\mathcal{L} &\supset& \frac{1}{2}\partial^\mu \varphi\partial_\mu \varphi \!-\! \frac{1}{2}m_\varphi^2\varphi^2 \!-\!\! \frac{g_{\varphi\gamma}}{4}\varphi\tilde{F}^{\mu\nu}F_{\mu\nu} \!-\! \frac{1}{4}F^{\mu\nu}F_{\mu\nu}\nonumber\\
	&& \!- i g_{\varphi\psi}\,\varphi\bar{\psi}\gamma^5\psi+i \bar{\psi} \gamma^\mu \left(\partial_\mu \!-\! iq A_\mu\right) \psi \!-\! m_\psi\bar{\psi}\psi,
\end{eqnarray}
where $F^{\mu\nu} = \partial^\mu A^\nu - \partial^\nu A^\mu$ is the field strength, $\tilde F^{\mu\nu}$ its dual, and we have introduced the Dirac matrices $\gamma^\mu$ and $\gamma^5$. Although the Lagrangian in Eq.~\eqref{eq:lagrangian} includes the interaction with fermions for completeness, this term can generally be neglected as justified in Appendix~\ref{app_A}.

The set of Maxwell's equations coupled to the axion field derived from Eq.~\eqref{eq:lagrangian} reads~\cite{Sikivie:1983ip, Wilczek:1987mv}
\begin{eqnarray}
	{\bf \nabla} \cdot {\bf E} &=& \rho_q - g_{\varphi\gamma}\left(\nabla\varphi\right)\cdot {\bf B}\,, \label{eq:gausslaw} \\
	{\bf \nabla} \cdot {\bf B} &=& 0\,,\label{eq:gausslawB}\\
	{\bf \nabla} \times {\bf E} &=& -\dot{\bf B}\,,\label{eq:faraday}\\
	{\bf \nabla} \times {\bf B} &=& {\bf J} + \dot{\bf E} + g_{\varphi\gamma}\dot{\varphi} {\bf B} + g_{\varphi\gamma}\left(\nabla\varphi\right)\times {\bf E}\,,\label{eq:ampere}
\end{eqnarray}
where the dot over the vector stands for a time derivative. The dynamics of the axion coupled to the electromagnetic field as derived from the Lagrangian is described by the Klein-Gordon equation,
\begin{equation}
	\label{eq_KG}
	\ddot{\varphi} - \nabla^2\varphi + m_\varphi^2\varphi= g_{\varphi\gamma}\,{\bf E}\cdot {\bf B}\,.
\end{equation}
Note, that the set of Maxwell's Eqs.~\eqref{eq:gausslaw}-\eqref{eq:ampere} in which the homogeneous equations are present is usually considered when investigating the coupling of the axion with the photon~\cite{Sikivie:1983ip, Wilczek:1987mv}. A different choice would consist in preserving the dual symmetry between the ${\bf E}$ and ${\bf B}$ fields, which would lead to the set in the dual axion representation explored in Ref.~\cite{Visinelli:2013fia}.


In the following, we decompose the equations above by choosing a reference frame $(\hat{\bf e}_x, \hat{\bf e}_y, \hat{\bf e}_z)$ and we break down the magnetic field as ${\bf B} = B_0 \hat{\bf e}_z + {\bf B}^{(1)}$, where the first term is a constant magnetic field along the axis $\hat{\bf e}_z$ and the second term is a perturbation which we write as
\begin{equation}
    {\bf B}^{(1)} = \int {\rm d}^3{\bf k}\,{\rm d}\omega\,e^{i{\bf k}\cdot {\bf x}-i\omega t}\!\!\!\sum_{s\in \{x, y, z\} }\!\!\!\hat{\bf e}_s\,\tilde{B}_s^{(1)}\,.
\end{equation}
A similar decomposition holds for the electric and the axion fields, as well for the sources $\rho_q$ and ${\bf J}$. We assume that the plasma wave propagates along the direction $\hat{\bf e}_z$ with wave number ${\bf k} = k \hat{\bf e}_z$, so we restrict ourselves to electrostatic waves only, for which ${\bf k} \parallel {\bf B}$. A different choice in which ${\bf k}$ has also a component along the, e.g.\ $\hat{\bf e}_x$ direction would lead to a different set of equations and to solutions that differ from the longitudinal modes explored here. See e.g.\ Refs.~\cite{Altherr:1990wi, Altherr:1990tf} for the general treatment.

With this choice, Faraday's Eq.~\eqref{eq:faraday} is decomposed as $k\tilde E_x = \omega \tilde B_y^{(1)}$, $k\tilde E_y = - \omega \tilde B_x^{(1)}$, and $\tilde B_z^{(1)} = 0$, so that Gauss's law for magnetism in Eq.~\eqref{eq:gausslawB} is also automatically satisfied. In addition, the last term in the modified Amp{\`e}re's Eq.~\eqref{eq:ampere} is negligible compared to the second to last because of Faraday's law. With this notation, the inhomogeneous Maxwell's equations read
\begin{eqnarray}
    \tilde E_z &=& \frac{1}{ik}\tilde \rho_q - gB_0\,\tilde \varphi\,,\label{eq:gauss1}\\
    k\tilde B_y^{(1)} &=& \omega\tilde E_x + i \tilde J_x \,,\\
    -k\tilde B_x^{(1)} &=& \omega\tilde E_y + i \tilde J_y \,,\\
    0 &=& \omega\tilde E_z +i \tilde J_z + g_{\varphi\gamma}B_0 \omega \tilde \varphi\,,\label{eq:faraday1}
\end{eqnarray}
so that both the components of the electric field that are orthogonal to the direction of wave propagation follow the expression $(\omega^2 - k^2) E_{x,y} = -i\omega J_{x,y}$. In fact, while the electromagnetic field in the directions orthogonal to propagation are unaffected by the modified equations, a transverse electric field component $E_z$ along the direction of propagation sourced by the axion field appears. For the moment, we leave aside Faraday's laws and we focus on Gauss' Eq.~\eqref{eq:gauss1} coupled with the Klein-Gordon Eq.~\eqref{eq_KG}, which is also modified as
\begin{equation}
	\label{eq_KG1}
	\left(-\omega^2 +k^2 + m_\varphi^2\right)\tilde\varphi= g_{\varphi\gamma}B_0\,\tilde E_z\,.
\end{equation}
Combining Eqs.~\eqref{eq:gauss1} and~\eqref{eq_KG1} leads to the expressions for the electric and axion fields,
\begin{eqnarray}
	\tilde{\varphi}(\omega, k) &=&\! -\frac{g_{\varphi\gamma}B_0}{ik}\frac{\tilde{\rho}_q(\omega, k)}{\omega^2 \!-\! k^2 \!-\! m_{\rm eff}^2}\,,\\
	\tilde E_z(\omega, k) &=&\! \left(1 \!+\! \frac{(g_{\varphi\gamma}B_0)^2}{\omega^2 \!-\! k^2 \!-\! m_{\rm eff}^2}\right) \!\frac{\tilde{\rho}_q(\omega, k)}{ik},\label{eq:tildeE}
\end{eqnarray}
where we introduced the effective mass squared $m_{\rm eff}^2 = m_\varphi^2 + (g_{\varphi\gamma}B_0)^2$. Faraday's Eq.~\eqref{eq:faraday1} is used to derive the expression for the current as $\tilde J_z = \tilde\rho_q \omega/k$, which in the space configuration can be restated as the continuity equation $\dot\rho + {\bf \nabla}\cdot{\bf J}=0$.

\section{Derivation of the axion-plasmon dispersion}
\label{dispersion}

The system of Maxwell and Klein-Gordon equations just introduced is closed once defined the phase-space distribution function $f_\lambda(t,{\bf x},{\bf p})$, where the subscript $\lambda$ stands for the charge species, either electrons ($e^-$), positrons ($e^+$), or ions ($I$), and ${\bf p}$ is the momentum of the species parcel. Such a distribution satisfies the collisionless Boltzmann equation
\begin{equation}
	\label{eq:Boltzmann}
	\left(\frac{\partial }{\partial t}+{\bf v}_\lambda\cdot {\bf \nabla} \right) f_\lambda + {\bf F}_\lambda\cdot {\bf \nabla}_{\bf p}f_\lambda=0\,,
\end{equation}
where ${\bf F}_\lambda$ is the force acting upon the species $\lambda$, ${\bf \nabla}_{\bf p}$ is the gradient in momentum space, ${\bf v}_\lambda = {\bf p} / E_\lambda$ is the parcel velocity, $E_\lambda = \sqrt{p^2+m_\lambda^2}$ is the energy for a particle of mass $m_\lambda$, and $p = |{\bf p}|$. For a charged plasma, particles are subject to the Lorentz force ${\bf F}_\lambda = q_\lambda \left({\bf E}+{\bf v}_\lambda\times {\bf B}\right)$, so that the dynamics of the phase-space distribution in Eq.~\eqref{eq:Boltzmann} reduces to the Vlasov equation with a negligible electron-axion coupling~\cite{Balakin:2013xha}.

For each species $\lambda$, we assume that the distribution fluctuates around an isotropic equilibrium configuration that depends only on $p$,
\begin{equation}
	\label{eq:expansion}
	f_\lambda(t, {\bf x}, {\bf p}) = f_{\lambda,0}(p)+ f_{\lambda,1}(t, {\bf x}, {\bf p})\,,
\end{equation}
with $f_{\lambda,1} \ll f_{\lambda,0}$ a perturbation. Assuming a quasi-neutral plasma implies that the equilibrium distributions satisfy $\sum_\lambda q_\lambda\,f_{\lambda,0} = 0$, so that the charge density is given by
\begin{equation}
    \label{eq:rho_micro}
	\rho_q\left(t, {\bf x}\right) = \sum_\lambda q_\lambda\,\int {\rm d}^3{\bf p}\, f_{\lambda,1}(t, {\bf x}, {\bf p})\,.
\end{equation}
Inserting the expansion of the phase-space density in Eq.~\eqref{eq:expansion} into Eq.~\eqref{eq:Boltzmann} and taking the Fourier transform of the resulting expression, the phase space perturbation is cast as
\begin{equation}
	\label{eq:vlasov_perturbations1}
	\tilde{f}_{\lambda,1}(\omega, k, {\bf p}) = -\frac{i\,q_\lambda}{\omega - {\bf k}\cdot {\bf v}_\lambda} \, \tilde{E}(\omega, k) \hat{\bf e}_z \cdot\frac{\partial f_{\lambda,0}({\bf p})}{\partial {\bf p}}\,.
\end{equation}
Finally, plugging Eq.~\eqref{eq:tildeE} into Eq.~\eqref{eq:vlasov_perturbations1}, multiplying by the charge $q_\lambda$, summing over the various species $\lambda$ and integrating over the momentum ${\rm d}p$ gives the plasma dispersion relation,
\begin{eqnarray}
	\label{eq:vlasov_perturbations2}
	&&G(\omega, k)\!\int \!\!\frac{{\rm d}^3{\bf p}}{k}\!\sum_\lambda\frac{\hat{\bf e}_z\cdot \hat{\bf p}}{{\bf k}\cdot {\bf v}_\lambda \!-\! \omega} q_\lambda^2 \frac{\partial f_{\lambda,0}}{\partial p} = 1\,,\\
	&&G(\omega, k) \equiv \frac{\omega^2-k^2-m_\varphi^2}{\omega^2-k^2-m_{\rm eff}^2}\,,
\end{eqnarray}
with $\hat{\bf p} \equiv {\bf p}/p$. In deriving Eq.~\eqref{eq:vlasov_perturbations2}, we have assumed that the equilibrium phase-space density depends on the magnitude of the momentum $p$ and not on its direction, as stated in the decomposition in Eq.~\eqref{eq:expansion}.

Once setting ${\rm d}^3{\bf p} = 2\pi p^2 {\rm d}p\,{\rm d}y$, where $y = {\bf \hat k}\cdot {\bf \hat p}$, the expression for the dispersion relation in Eq.~\eqref{eq:vlasov_perturbations2} reads
\begin{equation}
	\label{eq:vlasov_perturbations21}
	G(\omega, k)\!\int_0^{+\infty}\!\!\!{\rm d}p \int_{-1}^1{\rm d}y\frac{2\pi p^2}{k}\!\sum_\lambda\frac{y}{ykv_\lambda \!-\! \omega} q_\lambda^2 \frac{\partial f_{\lambda,0}}{\partial p} \!=\! 1\,,
\end{equation}
with $v_\lambda = p/E_\lambda$. We separate the real and imaginary parts of the complex integral by using the identity
\begin{equation}
	\frac{f(x)}{x-x_0} = {\rm PV} \frac{f(x)}{x-x_0} + i\pi \delta(x - x_0)\,,
\end{equation}
where $\delta(x)$ is the delta function. With this prescription, the integral over the angular direction in Eq.~\eqref{eq:vlasov_perturbations21} reads
\begin{eqnarray}
	\label{eq:def_Q}
	\int_{-1}^1 {\rm d}y\frac{y}{y - \omega/(kv_\lambda)} &=& 2 \!-\! \frac{\omega}{2kv_\lambda}\ln\left(\frac{\omega \!+\! kv_\lambda}{\omega \!-\! kv_\lambda}\right)^2 \\
	&& + i\pi\frac{\omega}{kv_\lambda}\Theta\left(v_\lambda - \frac{\omega}{k}\right)\,,\nonumber
\end{eqnarray}
where the Heaviside step function $\Theta(x)$ in the argument $x$ arises from the fact that for $\omega < kv_\lambda$ the integral in Eq.~\eqref{eq:def_Q} contains a pole. The imaginary part of the frequency in Eq.~\eqref{eq:vlasov_perturbations21} describes the damping of the wave. The step function implies that for $\omega > kv$ the imaginary component of the frequency is zero, while for $\omega < kv$ damping occurs. Integrating over the momentum, the step function yields to a lower integration limit at $p \equiv m_\lambda \bar u$, where $\bar u \equiv \omega/ \sqrt{k^2 - \omega^2}$. Finally, inserting Eq.~\eqref{eq:def_Q} into Eq.~\eqref{eq:vlasov_perturbations21} leads to the dispersion relation $\epsilon(\omega, k) = 0$, where the real and the imaginary components of the function $\epsilon(\omega, k)$ are respectively
\begin{eqnarray}
	\label{eq_dispersion_r}
    \epsilon_r(\omega, k)\! &=& 1+\frac{4\pi}{k^2}G(\omega, k)\sum_\lambda\int_0^{+\infty}{\rm d}p \frac{p^2}{v_\lambda}\times\\
    && \left[\frac{\omega}{4kv_\lambda}\ln\left(\frac{\omega + kv_\lambda}{\omega - kv_\lambda}\right)^2 - 1\right] q_\lambda^2 \frac{\partial f_{\lambda,0}}{\partial p}\,,\nonumber\\
    \epsilon_i(\omega, k)\! &=& \!- \frac{2\pi^2\omega}{k^3}\!G(\omega, k)\!\sum_\lambda\!\!\int_{m_\lambda\bar u}^{+\infty}\!\!\!\!{\rm d}p \frac{p^2}{v_\lambda^2}q_\lambda^2 \frac{\partial f_{\lambda,0}}{\partial p},	\label{eq_dispersion_i}
\end{eqnarray}
where each species $\lambda$ contributes to the branching of the solution on the $(\omega, k)$ plane, see e.g.\ Ref.~\cite{Stetina:2017ozh}. The condition for wave damping implies that all subluminal modes $\omega < k$ are damped, see Ref.~\cite{doi:10.1063/1.2353901}, while the imaginary part of the function in Eq.~\eqref{eq_dispersion_i} vanishes for $\omega \geq k$.

The choice of the notation for the function $\epsilon(\omega, k)$ recalls the definition of the dielectric function, for which the expressions in Eqs.~\eqref{eq_dispersion_r}--\eqref{eq_dispersion_i} are functionally identical, see e.g.\ Ref.~\cite{Silin:1960pya}. However, contrarily to the dielectric function which is a function of the real parameters $\omega$ and $k$, the frequency $\omega$ in the dispersion relation above is a complex number, with the imaginary component describing the damping of the wave. Note, that we introduce a small imaginary component in the frequency by setting $\omega \to \omega + i\gamma_L$, with $\gamma_L$ describing the absorption of the electrostatic and axion waves by the plasma. The presence of this imaginary term in the dispersion relation is linked to the phenomenon of Landau damping~\cite{LIFSHITZ1981217}. In fact, neglecting higher-order terms, the real ($r$) and imaginary ($i$) parts of the function $\epsilon(\omega, k)$ satisfy
\begin{equation}
	\epsilon(\omega+ i\gamma_L,k) \approx \epsilon_r(\omega,k)-i\frac{\partial \epsilon_r}{\partial \omega}\gamma_L + i\epsilon_i(\omega, k)\,,
	\label{eq_eps_decomp}
\end{equation}
so that the condition $\epsilon(\omega,k) = 0$ leads to
\begin{equation}
	\epsilon_r(\omega, k) = 0,\qquad \gamma_L = \left[\frac{\partial \epsilon_r(\omega,k)}{\partial \omega}\right]^{-1}\epsilon_i(\omega, k)\,.\label{eq:damping}
\end{equation}
As for the real component $\epsilon_r = \epsilon_r(\omega, k)$, the integration by parts of the expression in Eq.~\eqref{eq_dispersion_r} yields
\begin{eqnarray}
	\epsilon_r\!\! &=&\! 1 \!-\! \frac{4 \pi}{k^2}\,G(\omega, k)\sum_\lambda \int_0^{+\infty} {\rm d}p\frac{p^2}{E_\lambda(p)}\times\\
	&&\left(\!\frac{\omega}{kv_\lambda}\ln\frac{\omega \!+\! kv_\lambda}{\omega \!-\! kv_\lambda} \!-\! \frac{\omega^2 \!-\! k^2}{\omega^2 \!-\! k^2v_\lambda^2} \!-\! 1 \right)q_\lambda^2f_{\lambda,0}(p)\,,\nonumber
\end{eqnarray}
where the integral is related to the longitudinal polarization function $\Pi_L(\omega, k) \equiv \Pi^{00}(\omega, {\bf k})$ where $\Pi^{\mu\nu}$ is the electromagnetic polarization tensor, see Eq.(A17) in Ref.~\cite{Braaten:1993jw}.

\section{Results}
\label{results}

We now specialize Eq.~\eqref{eq_dispersion_r} to the case in which the plasma is composed of an electron-positron plasma, whose phase-space distribution at equilibrium is described by a Fermi-Dirac function~\cite{doi:10.1063/1.2353901}
\begin{eqnarray}
    f_{e^\pm}(p) &=& \frac{2}{(2\pi)^3}\frac{1}{e^{(\sqrt{p^2+m_e^2} \pm \mu)/T} + 1}\,,
\end{eqnarray}
where a plus (minus) sign denotes the positron (electron) distribution, $\mu$ is the electron chemical potential, and $m_e$ is the electron mass. Note, that the value of the chemical potential is fixed once the temperature and density of the plasma are given. Since electrons and positrons have the same mass and opposite charge, the expressions in the previous section simplify greatly and allow to define the number density of charged particles in the plasma as
\begin{eqnarray}
	\label{eq:normalize}
	n_0 &\equiv& 4\pi\int_0^{+\infty}{\rm d}p\,p^2\,\left[f_{e^+}(p) + f_{e^-}(p)\right]\\
    &=& \frac{1}{\pi^2} \int_0^{+\infty}{\rm d}p\,p^2\,\frac{e^{-\beta\Phi} + \cosh(\beta\tilde\mu)}{\cosh(\beta\Phi) + \cosh(\beta\tilde\mu)}\,,
\end{eqnarray}
where we introduced the notation $\beta \equiv m_e/T$, $\tilde \mu \equiv \mu/m_e$, and we switched to the variable $u = p/m_e$ so that we set $\Phi = \sqrt{u^2+1}$. The normalization in Eq.~\eqref{eq:normalize} acts as a definition of $\tilde\mu$ and has to be treated as a constraint over the expressions in Eqs.~\eqref{eq_dispersion_r}-\eqref{eq_dispersion_i}. Note, that the macroscopic current obtained from Faraday's Eq.~\eqref{eq:faraday1} and with the input from Eq.~\eqref{eq:rho_micro} leads to the microscopic description in terms of particle flux,
\begin{eqnarray}
    J_z &=& \frac{\omega}{k} 4 \pi e \int_0^{+\infty}{\rm d}p \,p^2 \left[f_{e^+}(p) - f_{e^-}(p)\right] \\
    &=& -\frac{\omega}{k} 4 \pi e \int_0^{+\infty}{\rm d}p\, p^2 \, \frac{\sinh(\beta\tilde\mu)}{\cosh(\beta\Phi) + \cosh(\beta\tilde\mu)}\,.\nonumber
\end{eqnarray}
Because of the continuity equation, a non-zero current density is related to a non-zero net charge of the plasma. We now consider an ultra-relativistic plasma, for which we neglect the chemical potential assuming charge neutrality.

\subsection{Ultrarelativistic plasma}

We first consider the case where the electron-positron plasma temperature is $T \gg m_e$, or $\beta \ll 1$, and with a negligible chemical potential. In this regime, the Fermi-Dirac distribution reduces to a Maxwell-Boltzmann distribution when the interparticle distance is larger than the thermal wavelength, $n_0^{1/3} \ll (2\pi m_eT)^{1/2}$. The expression in Eq.~\eqref{eq_dispersion_r} for the plasma dispersion relation with the phase-space distribution normalized as in Eq.~\eqref{eq:normalize} reads
\begin{eqnarray}
	\label{eq:vlasov_perturbations3}
	&&\frac{\omega^2-k^2-m_\varphi^2}{\omega^2-k^2-m_{\rm eff}^2}\!\left[\frac{\omega\,\beta}{4k\,K_2(\beta)}\mathcal{I}-1\right] = \frac{k^2}{2k_D^2},\\
	&&\mathcal{I} \equiv \int_0^{+\infty}\!\!\!\!\!\! {\rm d}u u \sqrt{1\!+\!u^2}\ln\!\left(\frac{\omega\sqrt{1+u^2} + ku}{\omega\sqrt{1\!+\!u^2} - ku}\!\right)^2\!e^{-\beta\sqrt{1+u^2}}\,,\nonumber
\end{eqnarray}
where the inverse Debye length for a plasma at temperature $T$ is
\begin{equation}
	\label{eq:debye_number}
	k_D = \left(\frac{4\pi\alpha_{\rm EM}}{T}n_0\right)^{1/2}\,.
\end{equation}
We first discuss the solutions of Eq.~\eqref{eq:vlasov_perturbations3} in the region $\omega \lesssim k$, for which the expression within the logarithmic term becomes independent on $u$ and the integral over momenta can be performed exactly using the normalization in Eq.~\eqref{eq:normalize}. In this regime, the dispersion relation is described by
\begin{equation}
	\label{eq_dispersion_rBUR}
	\frac{\omega^2 \!-\! k^2 \!-\! m_\varphi^2}{\omega^2 \!-\! k^2 \!-\! m_{\rm eff}^2}\left[\frac{\omega}{4k}\ln\left(\frac{\omega + k}{\omega - k}\right)^2 \!-\! 1\right]=\frac{k^2}{2k_D^2}\,,
\end{equation}
which, as expected for the ultrarelativistic case, is independent of the electron mass.

We separate the solutions into distinct regimes, and we introduce $k_\star^2 = 2k_D^2/3$ to facilitate the comparison with the results in Sec.~\ref{sec:degenerateplasma}. In the limit $k \ll \omega$, we expand the term in squared brackets in Eq.~\eqref{eq_dispersion_rBUR} as
\begin{equation}
    \label{eq:approx}
	\frac{\omega}{4kv}\ln\left(\frac{\omega + vk}{\omega - vk}\right)^2 -1 \approx \frac{v^2k^2}{3\omega^2} + \frac{v^4k^4}{5\omega^4} + \frac{v^6k^6}{7\omega^6} + ...\,,
\end{equation}
where $v= 1$ in the regime considered here, so that the dispersion relation can be obtained analytically. In the case $g_{\varphi\gamma} = 0$, the solution for the region $k \ll \omega$ is approximated in the region $k \lesssim k_\star$ by the relation~\cite{Silin:1960pya, Buti:1962}
\begin{equation}
	\label{eq:dispersion1}
	\omega^2 = k_\star^2 + \frac{3}{5}k^2 + \frac{12}{175}\frac{k^4}{k_\star^2}\,.
\end{equation}
In the same limit, the axion dispersion is given by the Klein-Gordon Eq.~\eqref{eq_KG1} as $\omega^2 = k^2 + m_\varphi^2$.

The inclusion of a non-zero axion-photon coupling affects these results by hybridizing the dispersion relations for the axion and the plasmon, effectively giving rise to an axion-plasmon polariton~\cite{Tercas:2018gxv}, with the two relations repelling each other by avoiding crossing. Inserting the expansion in Eq.~\eqref{eq:approx} to second order into Eq.~\eqref{eq_dispersion_rBUR} with $g_{\varphi\gamma} \neq 0$ leads to the dispersion relation for the upper and lower polariton modes,
\begin{equation}
    \label{eq:polaritons}
	\omega_{\rm U,L}^2 \!=\! \frac{1}{2}\!\left[k_\star^2 \!+\! k^2 \!+\! m_{\rm eff}^2 \pm \sqrt{\left(\!k_\star^2 - k^2 \!-\! m_{\rm eff}^2\right)^2 \!+\! 4 \Omega^4}\!\right].
\end{equation}
Here, $\Omega = \sqrt{g_{\varphi\gamma}B_0k_\star}$ is the analogous of the Rabi frequency, describing the strength of the axion-plasmon coupling~\cite{Tercas:2018gxv}. The solution for the upper polariton mode $\omega_{\rm U}$ is valid for a wide range of $k$ extending beyond $k \gtrsim k_\star$ and holds as long as the condition $k \ll \omega$ is satisfied. The lower polariton mode solution $\omega_{\rm L}$ is valid for $k \lesssim k_\ell$, with
\begin{equation}
	k_\ell = k_\star\,\frac{m_\varphi}{m_{\rm eff}}\,.
\end{equation}
The crossing from plasmon to axion behavior occurs at the wavelength $k = \sqrt{k_\star^2 - m_{\rm eff}^2}$.

A different solution for the lower polariton mode is provided in the region $k \lesssim \omega$. In fact, setting $\omega = k (1 + \delta)$ with an arbitrary $\delta$ into Eq.~\eqref{eq_dispersion_rBUR} yields
\begin{equation}
	\label{eq_dispersion_rBUR1}
	\frac{k^2\delta(2+\delta) \!-\! m_\varphi^2}{k^2\delta(2+\delta) \!-\! m_{\rm eff}^2}\left[\frac{1+\delta}{4}\ln\left(1+\frac{2}{\delta}\right)^2 \!-\! 1\right]=\frac{k^2}{2k_D^2}\,,
\end{equation}
As long as $|k^2 - \omega^2|\ll m_\varphi^2$, the limit $\delta \to 0$ leads to a solution in the region $k \lesssim \omega$ as
\begin{equation}
	\label{eq:dispersion22}
	\omega/k \!=\! 1 \!+\! 2\exp\left[-2\left(1 \!+\! \frac{k^2}{2k_D^2}\frac{m_{\rm eff}^2}{m_\varphi^2} \right)\right]\,.
\end{equation}
This relation is valid for $k_\ell \lesssim k \lesssim k_0$, where the momentum $k_0$ is obtained by setting $\epsilon_r = 0$ as
\begin{equation}
	\label{eq:vlasov_perturbations4}
	k_0 \equiv k_\ell\,\sqrt{2\ln(2/\beta)}\,.
\end{equation}
For $g_{\varphi\gamma}B_0 = 0$, the two expressions in Eqs.~\eqref{eq:dispersion1} and~\eqref{eq:dispersion22} cross at $k \approx 0.7\,k_D$. Note that, contrary to the solutions obtained in Eq.~\eqref{eq:polaritons} which is valid even in the case of a massless axion, the expression in Eq.~\eqref{eq:dispersion22} is not valid when discussing the region $k \sim \omega$ and for $m_\varphi = 0$ only the upper polariton solution $\omega_U$ exists. More in detail, Eq.~\eqref{eq_dispersion_rBUR1} in the presence of a massless axion does not have a solution for vanishing $\delta$, hence the dispersion relation does not cross the region $\omega \approx k$ and only solutions for a finite $\delta$ exist. To avoid this complication, in the following we consider a massive axion.

We finally compute the dispersion relation in the region $\omega \approx k$, where the phase velocity of the plasma waves approaches the speed of light~\cite{osti_4822425}. Assuming a linear approximation around $k \approx k_0$ for the dispersion of the kind $\Delta \omega = \Delta k - k_0\Delta v_p$, where the phase velocity is $v_p = \omega/k$, and taking into account the expression
\begin{equation}
	\frac{\Delta v_p}{\Delta k} = \frac{\partial \epsilon_r/\partial k}{\partial \epsilon_r/\partial v_p}\bigg|_{\omega = k}\,,
\end{equation}
with $\partial \epsilon_r/\partial v_p|_{\omega=k} = 12(k_\ell/\beta k_0)^2$ and $\partial \epsilon_r/\partial k|_{\omega=k} = 2/k_0$, we obtain the dispersion relation
\begin{equation}
	\label{eq:omegak}
	\omega = k - (k-k_0)\frac{\beta^2}{3}\ln(2/\beta)\,.
\end{equation}
The smallness of the term $(\beta^2/3)\ln(2/\beta)$ assures that the slope in Eq.~\eqref{eq:omegak} acts as a perturbation over the light cone solution $\omega = k$.

\begin{figure}[t!]
	\begin{center}
	\includegraphics[width=\linewidth]{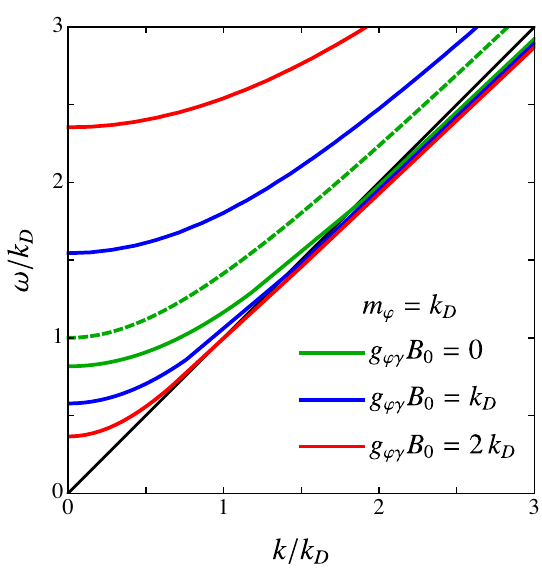}
	\caption{Kinetic dispersion relation of the axion-plasmon polariton modes for an ultrarelativistic plasma. The green dashed line is the axion dispersion for $g_{\varphi\gamma} = 0$. The plasma dispersion relation $\omega = \omega(k)$ for $\beta = 0.5$, for the illustrative case $m_\varphi = k_D$ and for three different values of the quantity $g_{\varphi\gamma} B_0 = 0$ (green solid line), $g_{\varphi\gamma} B_0 = k_D$ (blue solid line), and $g_{\varphi\gamma} B_0 = 2\,k_D$ (red solid line). Also shown for reference is the line $k = \omega$ (black solid line). All curves are in units of the inverse Debye length in Eq.~\eqref{eq:debye_number}.}
	\label{fig:PlotURma}
	\end{center}
\end{figure}

Results are shown in Fig.~\ref{fig:PlotURma}, where the frequency $\omega$ is plot as a function of the wavenumber $k$, both in units of $k_D$, and for the choices $\beta = 0.5$ and $m_\varphi = k_D$. In the case of no axion-photon coupling $g_{\varphi\gamma} = 0$, the plasma dispersion relation is described by the green solid line in Fig.~\ref{fig:PlotURma}, corresponding to Eq.~\eqref{eq_dispersion_rBUR} in the long-wavelength limit $k \ll \omega$ and approaching Eq.~\eqref{eq:omegak} for $k \approx \omega$. In this case, the dispersion relation of the axion field is represented by the green dashed line. When $g_{\varphi\gamma} \neq 0$, the upmost solution splits into the upper polariton mode $\omega_{\rm U}$ for $m_B = k_D$ (blue solid line) and $g_{\varphi\gamma}B_0 = 2k_D$ (red solid line), while the dispersion relation of the lower polariton mode, valid for $k \ll \omega$, is described by Eq.~\eqref{eq:dispersion22} for $k \lesssim \omega$ and by Eq.~\eqref{eq:omegak} for $k \approx \omega$. For reference, we also show the line $k = \omega$ (black solid line) to better visualize the deviations from the light-cone dispersion relation.

As discussed below Eq.~\eqref{eq:def_Q}, damping is effective for subluminal modes or, using Eq.~\eqref{eq:omegak}, for $k>k_0$. For $m_\varphi \gg g_{\varphi\gamma}B_0$, the result for $k_0$ in Eq.~\eqref{eq:vlasov_perturbations4} coincides with the usual expressions for the wave number describing Landau damping, while $k_0$ is suppressed by a factor $m_\varphi/(g_{\varphi\gamma}B_0)$ with respect to the usual results obtained in Ref.~\cite{Silin:1960} for $m_\varphi \ll g_{\varphi\gamma}B_0$. In this latter regime, the results obtained for the emission of axions in supernovae cores mediated by a plasmon would be suppressed by the same quantity~\cite{Mikheev:1998bg}.

To proceed with the computation of $\gamma_L$, we compute the derivative of $\epsilon_r$ with respect to $\omega$, using Eq.~\eqref{eq:omegak}. In this regime, the imaginary component of the function in Eq.~\eqref{eq_dispersion_i} reads
\begin{equation}
	\epsilon_i(\omega, k) = \frac{\pi\beta}{2}\frac{k_D^2}{k^2}\frac{m_\varphi^2}{m_{\rm eff}^2}\,\bar u^2e^{-\beta \bar u}\,.
	\label{eq_damping1}
\end{equation}
The last expression is valid for $\beta \ll 1 \ll \bar u$, which coincides with the ultrarelativistic limit and with the region $k \approx \omega$ where $\bar u \gg 1$ in which we are working. As the imaginary part of the function in Eq.~\eqref{eq_dispersion_i} is associated with the dissipation in the medium, the formula in Eq.~\eqref{eq_damping1} could seem counter-intuitive as it vanishes in the limit of a massless particle. However, in order to derive the dispersion relation in Eq.~\eqref{eq:omegak} we have also assumed that $m_\varphi \neq 0$; in this regards, one cannot take the limit $m_\varphi \to 0$ in Eq.~\eqref{eq_damping1}. For a massless axion, the dispersion relation in Eq.~\eqref{eq_dispersion_rBUR} does not cross the region $\omega = k$, so that the imaginary part of the function in Eq.~\eqref{eq_dispersion_i} is always vanishing. Finally, we recover the damping rate
\begin{equation}
	\label{eq:LandauDamping}
	\gamma_L(\Delta k) \approx \frac{\pi \beta k_0^2}{16\Delta k\ln(2/\beta)}e^{-\sqrt{\frac{3k_0}{\Delta k\ln(2/\beta)}}}\,,
\end{equation}
where $k_0$ has been defined in Eq.~\eqref{eq:vlasov_perturbations4}. Damping occurs only for the region $k_0+\Delta k$ with $\Delta k > 0$, namely for subluminal waves for which $\omega < k$. The function $\gamma_L(\Delta k)$ has a maximum for $\Delta k = (3/4)k_0/\ln(2/\beta)$ and it has decayed to a tenth of its maximum value for $\Delta k \approx 40\,k_0/\ln(2/\beta)$.

The results obtained are generally relevant for light axion in astrophysical setups. Consider for example the magnetized plasma surrounding a magnetar, in which the typical magnetic field is of the order of $B_0=10^{10}\,$T~\cite{Gunn:1969ej, Ostriker:1969if, Ferrari:1969}. If the effects studied here were not taken into account, the typical scale $k_0$ at which damping occurs would be expected to take place for $k_\ell \sim k_D\,\sqrt{\ln(2/\beta)}$ regardless of the axion mass. Instead, the inclusion of the axion-photon coupling leads to a suppression of the scale $k_0$ as in Eq.~\eqref{eq:vlasov_perturbations4}. For an axion-photon coupling in the range $g_{\varphi \gamma}\sim 10^{-10}{\rm\,GeV^{-1}}$, corresponding to the sensitivity of CAST~\cite{CAST:2017uph}, the suppression effect takes place for an axion mass $m_\varphi \lesssim g_{\varphi\gamma}B_0 \approx 1\,\mu$eV. Instead, if the axion-photon coupling lies below the projected sensitivity window of ADMX~\cite{ADMX:2021nhd}, $g_{\varphi \gamma}\lesssim 10^{-15}{\rm\,GeV^{-1}}$, the suppression is effective for the non-QCD axion masses $m_\varphi \lesssim g_{\varphi\gamma}B_0 \approx 10^{-11}\,$eV. In both scenarios, the dispersion relation is modified with respect to the standard case with $g_{\varphi\gamma} = 0$.

\subsection{Degenerate plasma}
\label{sec:degenerateplasma}

We now consider a highly degenerate Fermi gas, for which $T \ll \mu-m_e$. This is the limit to consider when dealing with the effects of oscillations in dense nuclear material such as the interior of neutron stars and white dwarfs, where the thermal length for the particles in the plasma is much larger than their interparticle distance. In this limit, the sum of the phase space distributions reads
\begin{equation}
    \label{eq:f0_deg}
    f_{e^+}(p) + f_{e^-}(p) = \frac{2}{(2\pi)^3}\Theta(p_F - p)\,,
\end{equation}
where $p_F = \sqrt{\mu^2-m_e^2}$ is the Fermi momentum and $\Theta(x)$ is the Heaviside function in the argument $x$, so that the normalization in Eq.~\eqref{eq:normalize} leads to the chemical potential $\mu = \sqrt{p_F^2+m_e^2}$ with $p_F = (3\pi^2n_0)^{1/3}$. Inserting Eq.~\eqref{eq:f0_deg} into Eq.~\eqref{eq_dispersion_r} leads to the dispersion relation
\begin{equation}
	\label{eq_dispersion_deg}
    \frac{\omega^2 \!-\! k^2 \!-\! m_\varphi^2}{\omega^2\!-\!k^2\!-\!m_{\rm eff}^2}\!\!\left[\!\frac{\omega}{4kv_F}\!\ln\!\left(\!\frac{\omega \!+\! kv_F}{\omega \!-\! kv_F}\right)^{\!\!2} \!\!-\! 1\!\right] \!\!=\! \frac{k^2}{2m_D^2},
\end{equation}
where the square of the Debye mass $m_D^2 \equiv (2\alpha_{\rm EM}/\pi)\mu^2v_F$ depends on the Fermi velocity $v_F \equiv p_F / \mu$, see e.g. Ref.~\cite{Stetina:2017ozh}. Note, that the dispersion relations for the degenerate regime in Eq.~\eqref{eq_dispersion_deg} has the same functional dependence on $\omega$ and $k$ as what has been previously obtained in the ultrarelativistic regime, see Eq.~\eqref{eq_dispersion_rBUR}. For this reason, the results in Eq.~\eqref{eq:polaritons} for $k \ll \omega$ are also valid for the degenerate case, once identified $k_\star^2 = 2m_D^2/3$. However, the Fermi velocity $v_F$ might not be close to the speed of light and appears in the solution even when the Fermi surface is nonrelativistic. Note, that when the axion-photon coupling is absent the modes satisfy ($v_p \equiv \omega/k$)
\begin{equation}
    \label{eq:requirement}
    \frac{v_p}{2v_F}\ln\left|\frac{v_p + v_F}{v_p - v_F}\right| > 1\,,
\end{equation}
or $\zeta k v_F \lesssim \omega$ with $\zeta \approx 0.833$. The appearance of the prefactor in Eq.~\eqref{eq_dispersion_deg} containing the axion-photon coupling leads to solution also for $\zeta k v_F \gtrsim \omega$ in the region $k^2 + m_\varphi^2 < \omega^2 < k^2 + m_\varphi^2 + (g_{\varphi\gamma}B_0)^2$.

For modes $k v_F \ll \omega$ the expansion of the terms in squared brackets in Eq.~\eqref{eq_dispersion_deg} to second order using Eq.~\eqref{eq:approx} gives the dispersion relation
\begin{equation}
	\label{eq_dispersion_deg1}
	\frac{\omega^2 - k^2 - m_\varphi^2}{\omega^2 - k^2 - m_{\rm eff}^2}\left(1 + \frac{3v_F^2k^2}{5\omega^2}\right) = \frac{\omega^2}{v_F^2k_\star^2}\,,
\end{equation}
so that for negligible couplings $m_\varphi \approx m_{\rm eff}$, we obtain the plasma dispersion $\omega^2 \approx v_F^2k_\star^2 + (3/5)v_F^2k^2$. Instead, when coupling cannot be neglected two separate solutions appear similar to what found in Eq.~\eqref{eq:polaritons} for the ultrarelativistic plasma.

\subsection{Nonrelativistic plasma}

The nonrelativistic case $\beta \gg 1$ has been extensively treated in Ref.~\cite{Tercas:2018gxv}. In this section, we show how the same results can be derived using the formalism discussed here. In the nonrelativistic regime, for which $\beta\,e^{-\beta\sqrt{1+q^2}}/K_2(\beta) \approx \sqrt{2/\pi} \beta^{3/2} e^{-\beta q^2/2}$, the logarithmic term in Eq.~\eqref{eq:vlasov_perturbations3} for modes whose phase velocity $\omega/k$ is much larger than the average velocity of particles in the plasma $v$ can be approximated as
\begin{equation}
	1 - \frac{\omega}{4kv}\ln\left(\frac{\omega + kv}{\omega - kv}\right)^2 \approx -\frac{v^2\,k^2}{3\omega^2}\,.
\end{equation}
In this approximation, the dispersion relation describing the upper- and lower-polariton modes reads
\begin{equation}
	\omega_{\rm U,L}^2 = \frac{1}{2} \left(\omega_{\rm Pl}^2 + \omega_\varphi^2 \pm \sqrt{\left(\omega_{\rm Pl}^2 - \omega_\varphi^2\right)^2 + 4 \Omega^4}\right)\,,
\end{equation}
where $\omega_{\rm Pl} \equiv (e^2 n_0/m_e)^{1/2} = k_D/\beta^{1/2}$ is the plasma frequency. This latter expression matches the results obtained in Ref.~\cite{Tercas:2018gxv} using the continuity and Navier-Stokes equations (respectively the zeroth and second moment of the Vlasov equation). In the absence of a photon-axion interaction, the two dispersion relations are $\omega^2 = \omega_{\rm Pl}^2$ and $\omega^2 = m_\varphi^2 + k^2$, corresponding respectively to the plasmon and axion modes. The Rabi frequency of the system $\Omega = \sqrt{\omega_{\rm Pl}\,g_{\varphi\gamma}B_0}$ describes two distinct effects, namely the mixing of the axion and plasmon modes and the repulsion between the upper and lower polariton modes. For both the upper and lower modes, the crossing from plasmon to axion behavior occurs when $\omega_{\rm Pl} = \omega_\varphi$, corresponding to the wavelength~\cite{Tercas:2018gxv}
\begin{equation}
	k_\star = \sqrt{\omega_{\rm Pl}^2 - m_{\rm eff}^2}\,.
\end{equation}

In the realm of the kinetic analysis performed here, each polariton mode acquires a small imaginary part as a consequence of the mixing of a Landau damped plasma mode, $\omega_{\rm U,L}\to \omega_{\rm U,L}+i\gamma_{\rm U,L}$, where the imaginary part reads
\begin{equation}
    \label{eq:damping2}
    \gamma_{\rm U,L} \approx -\sqrt{\frac{\pi}{8}}\frac{\omega_{\rm Pl}}{(k/k_D)^3}\!\left[1 \!+\! \frac{(g_{\varphi\gamma}B_0)^2}{\omega_{\rm U,L}^2 \!-\! k^2 \!-\! m_{\rm eff}^2}\right]\!e^{-\frac{\beta^2\omega_{\rm U,L}^2}{2k^2}}.
\end{equation}
For polariton frequencies close to the axion mode, $\omega_{\rm U,L}^2\sim k^2+m_{\rm eff}^2$, there is a resonant effect leading to effective axion Landau damping while, away from resonance, we obtain the plasma Landau damping rate, $\gamma_{\rm U,L}\approx \gamma_{\rm Pl}$, where~\cite{landau:book5}
\begin{equation}
\gamma_{\rm Pl} \simeq -\sqrt{\frac{\pi}{8}}\frac{\omega_{\rm Pl}}{(k/k_D)^3} e^{-\frac{k_D^2}{2k^2}}.    
\end{equation}
The results in Eqs.~\eqref{eq_damping1} and~\eqref{eq:damping2} may contain hints about the fate of axions at cosmological scales, namely in the conditions of  primordial and intergalactic plasmas. The QCD axion survives the strong magnetic fields in the radiation era, because of the large conductivity actually diminishes its dissipation in the medium and acts as a quantum Zeno effect~\cite{Ahonen:1995ky}. This could not be true in the hot intergalactic medium, which should have temperatures $T\sim 10^6\,$K and a density $n_e \sim 10^{-6}{\rm\,cm^{-3}}$, resulting in the plasma frequency $\omega_{\rm Pl} \sim 10^{-14}\,$eV. It is possible that particles of this mass do not survive the propagation in the intergalactic medium because of a much weaker magnetic field. These aspects will be treated with care in a separate publication.

\section{Summary and conclusions}
\label{conclusion}

We have derived the dispersion equations for the oscillations of a magnetized plasmas in the absence of collisions within the kinetic theory of a coherent superposition of axions and plasmons. The framework consists in the phase-space description of plasmas coupled to the axion field via the axion-photon term. In more detail, we construct a Vlasov$-$Maxwell$-$Klein-Gordon system in Eqs.~\eqref{eq:gausslaw}-\eqref{eq:ampere}, Eq.~\eqref{eq_KG}, and Eq.~\eqref{eq:Boltzmann}, capturing the key aspects of both relativistic and nonrelativistic plasmas: wave-wave and wave-particle interactions as well as the full relativistic nature of the axion field. In this regard, we substantially extend the focus of Ref.~\cite{Tercas:2018gxv} ans we set the stage for a set of future applications.

To illustrate some features of our kinetic theory, we have analytically obtained the dispersion relation of the axion-plasmon polaritons for both the ultrarelativistic and nonrelativistic plasmas, showing how the plasmon-axion interaction in the presence of a magnetic field modifies the plasma physics relations. We have obtained the dispersion relation for the plasma waves for the axion-polariton modes in the ultrarelativistic case and in different regimes. The dispersion relation split into an upper and a lower polariton modes due to the presence of the axion. Results are summarized in Fig.~\ref{fig:PlotURma} for a fixed axion mass $m_\varphi$ and for different values of the magnetic field $B_0$ and the axion-photon coupling $g_{\varphi\gamma}$. Landau damping is effective for wave numbers $k>k_0$, with $k_0$ depends on the ratio $m_\varphi/(g_{\varphi\gamma}B_0)$ and it might be suppressed in regions where $m_\varphi \ll g_{\varphi\gamma}B_0$. Similarly, damping is reduced in this regime as expressed by Eq.~\eqref{eq:LandauDamping}.

We have shown that Landau damping takes place in hot plasmas for some values of the plasma frequency, making possible for the axions to be kinematically damped in some situations of physical relevance. This latter, novel feature is worth being investigated in the context of astrophysical and cosmological plasmas, with potential to complement the literature on the mechanisms of axion production and suppression in dense and dilute plasmas. Moreover, we motivate novel experimental schemes to detect axions in laboratory plasmas in an active way, i.e.\ based on mechanisms where the axion is actually produced within the plasma, therefore complementing the running experiments based on telescopes and cavities, planned to detect axions produced at the interior of stars or in the primordial universe. 

Besides damping, the mixing of longitudinal modes with the axion also leads to other effects  including axion emission in the presence of a strong magnetic field~\cite{Mikheev:1998bg, Mikheev:2009zz, Caputo:2020quz}, acting both as a target for axion helioscopes and as a means to probe the inner solar structure. These effects depend on the relative strength of other competing terms entering the imaginary part of the Langmuir waves that also suffer all the other opacity sources such as Compton scattering or inverse bremsstrahlung.

\begin{acknowledgments}
LV acknowledges support by the Vetenskapsr\r{a}det (Swedish Research Council) through contract No.\ 638-2013-8993 and the Oskar Klein Centre for Cosmoparticle Physics. HT acknowledges Funda\c{c}\~{a}o da Ci\^{e}ncia e Tecnologia (FCT-Portugal) through Contract No. CEECIND/00401/2018, and through the Project No. PTDC/FIS-OUT/3882/2020. 
\end{acknowledgments}

\appendix

\section{Pauli equation with axion-electron interaction}
\label{app_A}

Here, we show under which conditions it is possible to neglect the axion-electron coupling for an electron in the presence of a magnetic field. Starting from the Lagrangian in Eq.~\eqref{eq:lagrangian}, the electron follows the equation of motion for the Dirac field
\begin{equation}
	i \gamma^\mu \left(\partial_\mu - iq A_\mu\right) \psi - m_e\psi + i g_{\varphi e}\varphi\gamma^5\psi = 0,
\end{equation}
where the coupling $g_{\varphi e} = C_e\,m_\varphi/f$, for a model-dependent factor $C_e$ of order one. We concentrate on the nonrelativistic case and we derive the expression analogous for the Pauli equation in the presence of the axion-electron coupling, writing the wave-function in terms of the large and small bi-spinor components as $\psi({\bf x}, t) = e^{-im_et}\left(\chi({\bf x}, t), \Phi({\bf x}, t)\right)$, where the time dependence of the bi-spinors occurs at scales larger than $1/m_e$. In the Dirac representation, we have
\begin{align}
	i \dot{\chi} \!&=\! -qA_0\chi \!-\! i g_{\varphi e}\varphi\Phi \!+\! i{\bf \sigma} \!\cdot\! {\bf \nabla} \Phi \!+\! q{\bf \sigma} \!\cdot\! {\bf A}\Phi,\label{eq:chi}\\
	i \dot{\Phi} \!&=\! -qA_0\Phi \!+\! i g_{\varphi e}\varphi\chi \!+\! i{\bf \sigma} \!\cdot\! {\bf \nabla} \chi \!+\! q{\bf \sigma} \!\cdot\! {\bf A}\chi \!-\! 2m_e\Phi \label{eq:Phi}.
\end{align}
Neglecting the time dependence of $\Phi$ and when $2m_e \gg qA_0$, Eq.~\eqref{eq:Phi} is rewritten as
\begin{equation}
	\label{eq:A4}
	\Phi = \frac{i}{2m_e}\left[g_{\varphi e}\varphi + {\bf \sigma}\cdot \left({\bf \nabla} -i q {\bf A}\right)\right]\chi.
\end{equation}
Substituting Eq.~\eqref{eq:A4} into Eq.~\eqref{eq:chi} and rearranging terms gives the Schr\"odinger equation in the presence of the axion-electron coupling,
\begin{equation}
	i\dot{\chi} \!=\! -qA_0 \chi \!-\! \frac{1}{2m_e}\!\left({\bf \nabla} \!-\! iq {\bf A} \right)^2\!\chi
	\!-\! \frac{\bf \sigma}{2m_e}\!\cdot\!\left(q {\bf B} \!+\! g_{\varphi e}\!{\bf \nabla} \varphi\right)\! \chi.
\end{equation}
The term $g_{\varphi e}{\bf \nabla} \varphi \sim m_\varphi/L$, where $L$ is the length over which the axion field varies, can be ignored with respect to $eB_0$ as long as
\begin{equation}
	L \gtrsim L_{\rm crit} \equiv 1{\rm \,fm} \left(\frac{B_0}{\rm T}\right)^{-1}\,\left(\frac{m_\varphi}{\rm \mu eV}\right).
\end{equation}
This condition is satisfied in many relevant astrophysical setups, such as: 1) Ultralight axions of mass $m_\varphi = 10^{-22}\,$eV that permeates the core of a DM halo of mass $\sim 10^{12}\,M_\odot$, so that de Broglie wavelength is $\lambda \approx 0.1\,$pc. A cloud of ultralight axions surrounding a supermassive black hole would also exhibit a large de Broglie wavelength of the order of an astronomical unit; 2) DM axions of mass $m_\varphi = 1\,\mu$eV and a typical galactic velocity dispersion $v = 10^{-3}c$, so that the de Broglie wavelength is $\lambda \approx 200\,$m.

\bibliographystyle{apsrev4-1}
\bibliography{axBib.bib}

\end{document}